\documentclass[aps,pra,twocolumn,notitlepage]{revtex4-1}

\usepackage{graphicx}
\begin{document}

\title{Contextual particle propagation in a three-path interferometer}

\author{Holger F. Hofmann}
\email{hofmann@hiroshima-u.ac.jp}
\affiliation{
Graduate School of Advanced Science and Engineering, Hiroshima University,
Kagamiyama 1-3-1, Higashi Hiroshima 739-8530, Japan}

\begin{abstract}
Quantum information is based on the apparent contradictions between classical logic and quantum coherence described by Kochen-Specker contextuality. Surprisingly, this contradiction can be demonstrated in a comparatively simple three-path interferometer, where it is impossible to trace the path of a single photon through five consecutive stages of the interferometer. Here, I discuss the paradoxical aspects of single photon interferences revealed by the three-path interferometer and point out the essential role of dynamics in quantum information. 
\end{abstract}

\keywords{quantum measurement, quantum contextuality, quantum interference}

\maketitle

\section{Introduction}

Quantum information theory can be traced back to very early attempts to reconcile Hilbert space relations with Boolean logic \cite{Bir36}. Since logic is fundamentally concerned with statements that can be true or false, the starting point is a set of logical propositions that are identified with measurement outcomes. This identification is problematic because the different measurements are not compatible with each other. It is not at all clear why a specific outcome of measurement x is represented by a mathematical ``superposition'' of the outcomes of a measurement y. The consequences of this dilemma are encapsulated in quantum contextuality as expressed by the Kochen-Spekker theorem \cite{KS}. Essentially, the theorem shows that the components of Hilbert space vectors can only be identified with the outcomes of a measurement when the specific measurement is actually performed. 

It should be obvious that this situation represents a dilemma. We can either try to force quantum theory into a procrustean bed of mathematical logic by looking for standardized sets of measurements, or we can look for a better explanation of the physics described by the quantum formalism. The former approach has motivated the recent generalization of contextuality to arbitrary combinations of state preparation and measurement \cite{Spe05,Spe08,Pus14,Sch18}, which seems to indicate that contextuality is best expressed in terms of negativity in the weak values of projectors \cite{Aha88,Lei05,Tol07}. The latter is consistent with the observation that complex weak values provide the most consistent statistical description of quantum correlations between observables that cannot be measured at the same time \cite{Hof12,Hof14}. Despite these important discoveries, it seem that the opportunities for a better explanation of quantum physics are being overlooked by researchers committed to a one-sided approach based on logical propositions. It is therefore more than urgent to put the physics back into quantum information. 

In our recent research, we have shown that weak values describe fundamental features of quantum dynamics and that the incompatibility between different measurements arises from these dynamics \cite{Hof11,Pat19,Mat21,Mat23}. We have also re-formulated Kochen-Specker contextuality in a way that allows us to identify the unitary transformations between contexts as the origin of the non-classical relation between different statistics \cite{Ji23} and demonstrated the accuracy of a weak value description of quantum interferences at the level of individual particles \cite{Hof21,Lem22}. All of these results strongly support the idea that quantum contextuality is best understood in terms of the quantum interference effects observed in single particle interferometers. In order to demonstrate this point, I have recently introduced a three-path interferometer in which the interfering paths represent the five different measurement contexts needed for a demonstration of contextuality in a three dimensional Hilbert space \cite{Hof23a}. Here, I will summarize the general features of the three-path interferometer and illustrate its application to an input state with a particularly strong violation of the inequality associated with non-contextual hidden variable models. The specific focus in this presentation is on the consistency of the conditional currents described by weak values of the path projectors. The particular choice of input state discussed here highlights the fact that negative conditional currents represent a conditional delocalization of individual particles, as previously demonstrated in the context of feedback compensation scenarios \cite{Hof21,Lem22,Hof23b}.

\section{The three-path interferometer}

\begin{figure}
\begin{picture}(240,120)
%%\put(0,0){\framebox(240,120){}}
\put(0,0){\makebox(240,120){\vspace{-2cm}
\scalebox{0.6}[0.6]{
\includegraphics{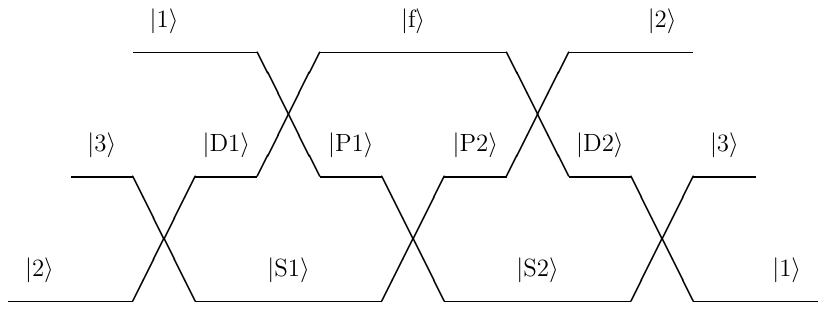}}}}
\end{picture}

\caption{\label{fig1}  
Implementation of contextuality in the three-path interferometer \cite{Hof23a}. Parallel paths are represented by orthogonal states. Each beam splitter represents a transformation of the path-basis into a new measurement context. 
}

\end{figure}

The three-path interferometer introduced in \cite{Hof23a} gives a specific physical meaning to five different measurement contexts, where two subsequent contexts always share one of their three measurement outcomes. In an interferometer, the transition from one context to the next is represented by beam splitters that interfere two paths of the input context to produce the corresponding two paths of the output contexts. The path that is shared by the two contexts remains parallel to the two contexts. Similar to a Mach-Zehnder interferometer, the output paths are identical to the input paths when no phase shifts are applied. Hence photons entering the interferometer in the input paths $\{1,2,3\}$ exit the interferometer in the corresponding output paths $\{1,2,3\}$. The second context is given by the paths $\{1,S1,D1\}$, where a beam splitter of reflectivity $R_1=1/2$ implements the relations
\begin{eqnarray}
\mid S1 \rangle &=& \frac{1}{\sqrt{2}} \left( \mid 2 \rangle + \mid 3 \rangle\right)
\nonumber \\
\mid D1 \rangle &=& \frac{1}{\sqrt{2}} \left( \mid 2 \rangle - \mid 3 \rangle\right).
\end{eqnarray}
The third context $\{S1,f,P1\}$ shares the path $S1$ with the second context and the paths $D1$ and $1$ interfere at a beam splitter of reflectivity $R_{S1}=1/3$, implementing
\begin{eqnarray}
\mid f \rangle &=& \frac{1}{\sqrt{3}} \left( \mid 1 \rangle + \mid 2 \rangle - \mid 3 \rangle\right)
\nonumber \\
\mid P1 \rangle &=& \frac{1}{\sqrt{6}} \left( 2 \mid 1 \rangle - \mid 2 \rangle + \mid 3 \rangle\right).
\end{eqnarray}
To achieve perfect symmetry, the fourth context $\{S2,f,P2\}$ is implemented by a beam splitter with reflectivity $R_f=1/4$, so that 
\begin{eqnarray}
\mid P2 \rangle &=& \frac{1}{\sqrt{6}} \left( - \mid 1 \rangle + 2 \mid 2 \rangle + \mid 3 \rangle\right)
\nonumber \\
\mid S2 \rangle &=& \frac{1}{\sqrt{2}} \left( \mid 1 \rangle + \mid 3 \rangle\right).
\end{eqnarray}
The fifth and final context $\{2,S2,D2\}$ is obtained by interference between path $f$ and path $P2$ at a beam splitter of reflectivity $R_{S2}=1/3$, so that 
\begin{equation}
\mid D2 \rangle = \frac{1}{\sqrt{2}}\left(\mid 1 \rangle - \mid 3 \rangle\right).
\end{equation}
The final beam splitter of reflectivity $R_2=1/2$ restores the original context $\{1,2,3\}$. The correspondence between interferometer paths and quantum states is shown in Fig. \ref{fig1}. 

If photons propagate through the interferometer as classical particles, each photon must travel along a well-defined sequence of paths. We can then consider photons that pass through path $f$ on their way from input to output. For each input path, there is only one way to get to the correct output via $f$. For input path $1$, this is $1-f-D2-1$, for input path $2$, it is $2-D1-f-2$, and for $3$ it is $3-D1-f-D2-3$. Each of the options includes either $D1$ ($2$) or $D2$ ($1$) or both ($3$). One would therefore conclude that the probability of finding the photon in path $f$ could not be higher than the sum of the probabilities of finding the photon in $D1$ or $D2$,
\begin{equation}
    P(f) \leq P(D1) + P(D2).
\end{equation}
This is the widely known inequality violated by quantum contextuality in three dimensional Hilbert spaces in the notation introduced in \cite{Ji23} and applied to the three-path interferometer in \cite{Hof23a}.  

\section{Contextuality in the path probabilities}
\begin{figure}
\begin{picture}(240,120)
%%\put(0,0){\framebox(240,120){}}
\put(0,0){\makebox(240,120){\vspace{-2cm}
\scalebox{0.6}[0.6]{
\includegraphics{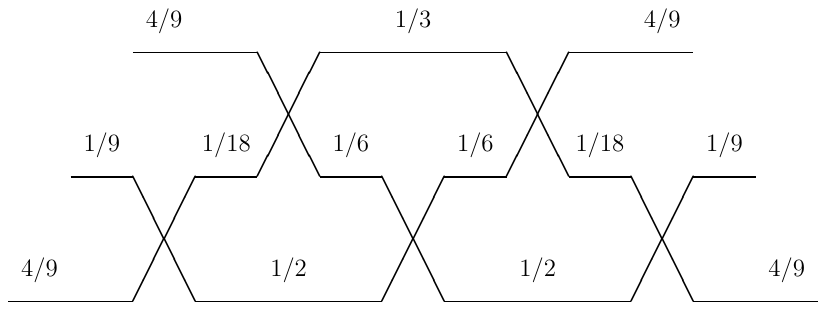}}}}
\end{picture}

\caption{\label{fig2}  
Probabilities of photon detections in the paths of the interferometer for the state $\mid N_x \rangle$. The violation of the inequality is apparent in the low probabilities of $P(D1)=P(D2)=1/18$ and the high probability of $P(f)=2/9$.  
}
\end{figure}

A wide range of different quantum states can violate the inequality in Eq.(\ref{eq:ineq}). Most works on contextuality focus on the case of $P(D1)=P(D2)=0$, a well defined pure state input that violates the inequality since it has a probability of $P(f)=1/9$ in the path $f$. As discussed in \cite{Hof23a}, it then seems as if photons in $f$ could only flow from input $1$ to input $2$, in direct contradiction of the propagation of amplitudes through the setup. Here, I want to take the opportunity to look at a state with non-zero probabilities in $P(D1)$ and $P(D2)$, where $P(f)$ can be significantly higher than $1/9$. The state in question is given by the input superposition
\begin{equation}
\mid N_x \rangle = \frac{1}{3}\left( 2\mid 1 \rangle + 2 \mid 2 \rangle + \mid 3 \rangle\right).
\end{equation}
The probabilities of finding a single photon in the different paths of the interferometer are shown in Fig. \ref{fig2}.

The violation of the inequality in Eq.(\ref{eq:ineq}) is expressed by the three probabilities 
\begin{eqnarray}
    P(f) &=& |\langle f \mid N_x \rangle|^2 = 1/3 \nonumber \\
    P(D1) &=& |\langle D1 \mid N_x \rangle|^2 = 1/18 \nonumber \\
    P(D2) &=& |\langle D2 \mid N_x \rangle|^2 = 1/18.
\end{eqnarray}
The inequality is violated by a difference of 
\begin{equation}
P(f)-P(D1)-P(D2)=2/9.
\end{equation}
If photons propagated as classical particles, this fraction of the total photon number would have to propagate through $f$ by passing from $1$ to $2$.  

\section{Consistency of conditional currents}
As argued in \cite{Hof23a}, a more consistent explanation of photon propagation in the three-path interferometer makes use of the conditional currents defined by the weak values of the path projectors. This explanation is not only more consistent with the propagation of classical waves through the interferometer, it can also be confirmed in experiments using feedback compensation \cite{Hof21,Lem22,Hof23b}. The current through $i$ conditioned by an output $o$ is given by the weak value
\begin{equation}
W(i|o) = \frac{\langle o \mid i \rangle \langle i \mid N_x \rangle}{\langle o \mid N_x \rangle}.
\end{equation}
The total current through $f$ is given by
\begin{equation}
P(f) = W(f|1) P(1) + W(f|2) P(2)+ W(f|3) P(3), 
\end{equation}
with
\begin{eqnarray}
    W(f|1)&=&1/2 \nonumber \\
    W(f|2)&=&1/2 \nonumber \\
    W(f|3)&=&-1.
\end{eqnarray}
%%%%
\begin{figure}
\begin{picture}(240,120)
%%\put(0,0){\framebox(240,120){}}
\put(0,0){\makebox(240,120){\vspace{-2cm}
\scalebox{0.6}[0.6]{
\includegraphics{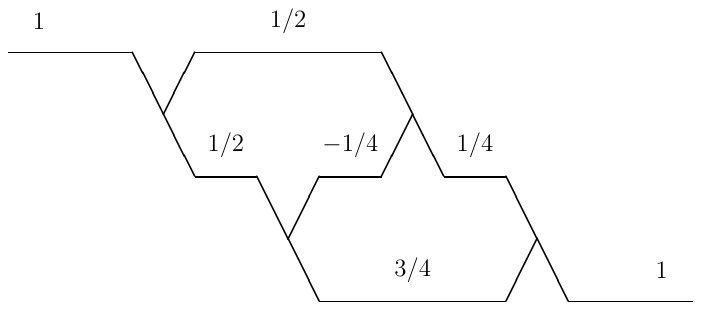}}}}
\end{picture}
\caption{\label{fig3}  
Conditional currents from input $1$ to output $1$. The positive current through $f$ merges with the negative current through $P2$, satisfying the continuity relation with the current through $D2$.
}
\end{figure}
%%%%
The conditional currents in $D2$ are
\begin{equation}
P(D2) = W(D2|1) P(1) + W(D2|3) P(3) 
\end{equation}
with 
\begin{eqnarray}
    W(D2|1)&=&1/4 \nonumber \\
    W(D2|3)&=&-1/2.
\end{eqnarray}
Continuity requires that the conditional currents flowing into a beam splitter must also exit the beam splitter. In the case of currents conditioned by $1$,
\begin{equation}
W(D2|1) = W(f|1)+W(P2|1).
\end{equation}
The reason why the conditional current in $D2$ is lower than the conditional current in $f$ is the negative conditional current in $P2$,
\begin{equation}
W(P2|1) = -1/4.   
\end{equation}
Fig. \ref{fig3} illustrates this continuity of the currents conditioned by the detected outcome $1$. 

Negative conditional currents provide a microscopic description of quantum contextuality. The assignment of continuous and possibly negative currents to the measurement outcomes $(1,2,3)$ indicates that the reality of finding a photon in any other path $\mid i \rangle$ depends on the physical implementation of that measurement. Each measurement thus assigns a discrete reality to the actually detected paths and a continuous current to the paths along which the undetected photon travelled. 

\section{Conclusions}

The propagation of a single photon from an unknown input port to a corresponding output port provides an example of quantum dynamics that can be formulated in terms of quantum information concepts due to the discrete nature of detection events in the paths. The problem of contextuality then corresponds to the problem of interferences between context independent realities. The continuity of currents through the interferometer and the consistency between weak values and interference effects then suggests that quantum superpositions should not be interpreted in terms of hypothetical measurement outcomes when these outcomes are not actually obtained in the physical situation that is being considered. Instead, it seems necessary to consider contextual continuity as an essential feature that distinguishes classical logic from quantum information.

%%\vfill

\end{document}